\documentclass[a4paper, 11pt]{article}

\usepackage{lmodern}
\usepackage[T1]{fontenc}
\usepackage[latin9]{inputenc}

\pdfoutput = 1

\usepackage{styleBuding}
\usepackage{bm}
\usepackage{color,listings}
\usepackage{amsthm, amsmath}
\usepackage{amssymb}
\usepackage{mathrsfs}
\usepackage{graphicx}
\usepackage{slashed}
\usepackage{soul}
\usepackage[dvipsnames]{xcolor}
\usepackage{float}
\usepackage{xurl,hyperref}
\usepackage{enumitem}
\hypersetup{colorlinks=true,citecolor=red,linkcolor=NavyBlue,urlcolor=NavyBlue}
\usepackage{multirow}
\usepackage{subfigure}
\usepackage{diagbox}
\usepackage{url} 
\usepackage[figuresright]{rotating}
\definecolor{back}{HTML}{F8F8F8}
\usepackage{epstopdf}
\usepackage{lipsum}

\usepackage{booktabs}  
\usepackage{array}

\newcommand{\rom}[1]{\uppercase\expandafter{\romannumeral #1\relax}}

\allowdisplaybreaks
\DeclareUnicodeCharacter{2212}{-}

\title{Constraining ALP-Top Interaction from the Chromoelectric Dipole Moment of the Top Quark}
\author{Subhadip Bisal$^{a,b}$}
\affiliation{$^a$ School of Physical Sciences, Indian Association for the Cultivation of Science,
2A $\&$ 2B, Raja S.C. Mullick Road, Jadavpur, Kolkata 700032, India}
\affiliation{$^b$ School of Physics, Zhengzhou University, Zhengzhou 450000, China}
\emailAdd{subhadipbisal6@gmail.com}

\abstract{The couplings of axion-like particles (ALPs) to Standard Model fermions are proportional to the fermion masses, making the interaction with the top quark particularly significant. In this study, we consider an ALP that is a mixture of CP-even and CP-odd components, thereby introducing CP violation. This CP violation, in turn, gives rise to electric dipole moments (EDMs) of quarks and leptons, as well as chromoelectric dipole moments (CEDMs) of quarks. We compute the one-loop and two-loop contributions to the top quark CEDM induced by the ALP. In our calculation, we treat the external gluon as off-shell with momentum $q^2 \neq 0$, derive the analytical results, and finally evaluate the top quark CEDM at $q^2 = m_t^2$, corresponding to the top quark pole mass. This value is relevant for subsequent calculations of the EDMs of the neutron and mercury. By applying current experimental bounds on EDMs and CEDMs, we derive constraints on the ALP-top quark coupling, with the strongest limit coming from the neutron EDM.}

\begin{document}

  \maketitle
  \flushbottom

\section{Introduction}
The lack of discoveries related to heavy new physics at the Large Hadron Collider (LHC), which was largely anticipated due to the weak-scale hierarchy problem, has encouraged a change in direction. As a result, scenarios featuring light mediators are gaining growing attention in both theoretical studies and experimental searches. Light spin-0 particles, whether scalar, pseudoscalar or a combination of both, are often collectively known as axion-like particles (ALPs). These types of particles frequently arise in theories beyond the Standard Model (SM). Their small mass, when compared to the typical scale of new physics, can often be attributed to their nature as pseudo-Nambu-Goldstone bosons, meaning they are associated with an underlying broken symmetry.  ALPs represent a broader class of particles that includes the well-known QCD axion, but differ in that their mass and couplings are free parameters, subject to experimental constraints. Interestingly, ALPs offer possible solutions to several fundamental questions in particle physics. These include the strong CP problem~\cite{Peccei:1977hh, Peccei:1977ur, Weinberg:1977ma, Wilczek:1977pj}, the nature and origin of dark matter~\cite{Preskill:1982cy, Abbott:1982af, Dine:1982ah, Davis:1986xc}, and the unresolved issues related to flavor~\cite{Davidson:1981zd, Wilczek:1982rv, Ema:2016ops, Calibbi:2016hwq} and mass hierarchies~\cite{Graham:2015cka}.

The search for ALPs with masses below the MeV scale has spurred a wide array of experimental efforts, many of which intersect with astrophysical and cosmological studies~\cite{Jaeckel:2010ni, Marsh:2015xka, Irastorza:2018dyq, DiLuzio:2020wdo}. 
In the sub-eV range, experiments employ \textit{wave-based} detection techniques such as haloscopes, helioscopes, and optical setups, while beam-dump experiments extend sensitivity up to the GeV scale~\cite{Dobrich:2019dxc}. 
At higher energies, collider experiments have probed ALP masses from the GeV level to the electroweak (EW) scale by analyzing their associated production with photons, jets, and EW bosons~\cite{Jaeckel:2015jla, Knapen:2016moh, Brivio:2017ije, Mariotti:2017vtv, CidVidal:2018blh, Aloni:2018vki, Aloni:2019ruo, Baldenegro:2018hng, Barbosa:2025zyn}. 
Furthermore, searches for exotic on-shell $Z$ boson and Higgs decays into ALPs have revealed previously unexplored regions of the parameter space~\cite{Bauer:2017ris, Bauer:2017nlg}.

Interestingly, CP-violating signatures of ALPs have received relatively little attention so far~\cite{Marciano:2016yhf, Moody:1984ba, Pospelov:1997uv, Raffelt:2012sp, Bertolini:2020hjc, OHare:2020wah, Stadnik:2017hpa, Dzuba:2018anu}. 
The CP symmetry is violated when the ALP ($a$) couples to fermions ($\psi_f$) via a combination of scalar or CP-even $(a \bar{\psi}_f \psi_f)$ and pseudoscalar or CP-odd $(a \bar{\psi}_f \gamma_5 \psi_f)$ operators.\footnote{The CP symmetry is also broken when the ALP-photon coupling includes both $aF_{\mu\nu}\tilde{F}^{\mu\nu}$ and $aF_{\mu\nu}F^{\mu\nu}$ interactions, where $F^{\mu\nu}$ is the electromagnetic field strength tensor and $\tilde{F}^{\mu\nu}$ denotes its dual.} 
If ALPs exhibit CP-violating interactions, they could induce fermion electric dipole moments (EDMs)~\cite{Marciano:2016yhf, DiLuzio:2020oah}. 
Similar to leptons and quarks possessing an EDM, quarks can also acquire chromoelectric dipole moments (CEDMs) through their interactions with external gluon fields. 

The top quark holds a unique position in the study of the SM phenomenology. This is because its mass is approximately the same as the EW symmetry breaking (EWSB) scale. Unlike other quarks, the top quark is unique because its very short lifetime ($5\times 10^{-25}$~sec) prevents it from hadronizing. Instead, it decays semi-weakly and has a Yukawa coupling of $\mathcal{O}(1)$. At a center-of-mass (c.o.m) energy of $\sqrt{s}=14$~TeV, approximately 90\% of top quark production results from gluon fusion ($gg\to t\bar{t}$), with the remaining fraction comes from $q\bar{q}$ annihilation~\cite{LHCHiggsCrossSectionWorkingGroup:2013rie}. Therefore, unlike the lepton EDM, which is measured in low-energy experiments (e.g., atomic/molecular spectroscopy), the top quark CEDM, however, can only be measured dynamically in high-energy collisions.
In many beyond the SM (BSM) scenarios, the CEDM has been studied at the one-loop level in the presence of CP-violating interactions (see, e.g., Refs.~\cite{Huang:1994zg, Ibrahim:2011im, Aboubrahim:2015zpa, Gorbahn:2014sha, Hernandez-Juarez:2018uow, Hernandez-Juarez:2020gxp, DiLuzio:2020oah, Gisbert:2021htg, Aiko:2025tbk}). Similarly, two-loop contributions to the CEDM have also been studied in various BSM contexts (see, e.g., Refs.~\cite{Abe:2013qla, Chien:2015xha, Nakai:2016atk, DiLuzio:2020oah, Gisbert:2021htg, Aiko:2025tbk}). Related studies on ALP-induced EDM and chromomagnetic dipole moment (CMDM) are discussed in Ref.~\cite{Neubert:2024jal}.

The current experimental upper limit on the CEDM of top quark is~\cite{CMS:2019kzp} 
\begin{align}
    \left\lvert \hat{d}_t^C\right\rvert_{\rm Exp.} < 0.03 \; \; \text{at $95\%$ confidence level (C.L.)}. 
    \label{eq:cedmbound}
\end{align}

This constraint is derived from measurements of the CMDM of top quark by the CMS collaboration at the LHC, using $pp$ collisions with a c.o.m energy of 13~TeV and an integrated luminosity of 35.9~fb$^{-1}$~\cite{CMS:2019kzp}.

In this study, we calculate the CEDM of the top quark mediated by an ALP. Focusing on CP-violating ALP-top quark interactions, we demonstrate how these induce a CEDM. We perform computations at both one-loop and two-loop levels, with the two-loop contribution originating from a Barr-Zee-type diagram.
As has already been stated, the CEDM of the top quark can only be measured at high energies, we account for this by considering the external gluon to be off-shell, i.e., carrying a nonzero momentum transfer ($q^2 \neq 0$). We perform a complete two-loop computation and present the analytical formulae for the form factors contributing to the top quark CEDM for $q^2 \neq 0$. We then evaluate the numerical values at the top quark pole mass, i.e., at $q^2 = m_t^2$. To the best of my knowledge, this result is not available in the existing literature. We further extend our analysis to determine the contribution of the top quark CEDM to the Weinberg operator, as well as to the EDMs and CEDMs of light quarks, which are induced via the top CEDM. Using these results, we then compute the EDMs of the neutron and mercury, both of which crucially depend on the top quark CEDM evaluated at its pole mass. Finally, employing the current experimental bounds on the top quark CEDM (see Eq.~\eqref{eq:cedmbound}) and the limits on the neutron and mercury EDMs, we derive constraints on the ALP-top coupling.


The paper is structured as follows. In Sec.~\ref{sec:alps}, we provide a brief introduction to the CP-violating ALPs and their couplings to fermions. Sec.~\ref{sec:CMDMintro} discusses the CEDM, EDM, and the Weinberg operator within the effective Lagrangian framework. We then present the one-loop calculation and a detailed two-loop calculation of the top quark CEDM mediated by the ALP in Sec.~\ref{sec:twoloopcal}, maintaining full generality. In Sec.~\ref{sec:edmcedmweinberg}, we analyze the impact of the top quark CEDM on the Weinberg operator, as well as on the neutron and mercury EDMs. Sec.~\ref{sec:limitothersearches} summarizes existing constraints on the ALP-top coupling from collider searches (CMS, ATLAS) and low-energy precision experiments (NA62, BABAR). Our main results, including limits on the ALP-top coupling derived from top quark CEDM, neutron EDM, and mercury EDM, are presented in Sec.~\ref{sec:mainresults}. Finally, we conclude with a summary of our findings in Sec.~\ref{sec:conclusion}.

\section{The CP-Violating ALPs}
\label{sec:alps}

The ALP exhibits a broad spectrum of possible couplings. In this work, we focus exclusively on the standard CP-even and CP-odd interactions between the ALP, denoted as $a$, and fermions $\psi_f$, which induce CP-violating effects. These interactions give rise to EDMs of leptons and quarks, as well as CEDMs of quarks. Additional couplings, such as those between the ALP and gauge bosons, can be generated at loop level.
The effective Lagrangian describing the ALP-fermion interaction takes the following form~\cite{Hisano:2012cc, Maltoni:2024wyh, DiLuzio:2020oah, Marciano:2016yhf, OHare:2020wah, DiLuzio:2023lmd, DiLuzio:2023cuk}:
\begin{align}
    \mathcal{L}^a =-  \sum_{\psi_f}^{} \frac{m_f}{f_a}c_f\, a\bar{\psi}_f \psi_f -  \sum_{\psi_f}^{} i \frac{m_f}{f_a}\tilde{c}_f \, a\bar{\psi}_f \gamma_5 \psi_f~,
    \label{eq:ALPint}
\end{align}
where $m_f$ represents the mass of the fermion $f$, while $c_f$ and $\tilde{c}_f$ are the real parameters. The quantity $f_a$ corresponds to a heavy new scale, with $f_a\gg v$, where $v=246~\text{GeV}$ denotes the EW scale. In Eq.~\eqref{eq:ALPint}, the first term is CP-even while the second term is CP-odd. Therefore, the combination of these terms violates CP.  
The pseudoscalar interactions (CP-odd) can be expressed in a shift-symmetric form using the dimension-five operator $\sim \frac{\partial_\mu a}{2f_a}(\bar{f}\gamma^\mu \gamma_5 f)$, derived through integration by parts and application of the equations of motion. This formulation naturally leads to the normalization factor $\frac{m_f}{f_a}$. In contrast, the scalar interactions (CP-even) explicitly break the shift symmetry. The scalar interactions in the unbroken phase of the SM can be described by the dimension-five operator $aH\bar{\psi}_{f,\,L(R)}\psi_{f,\,R(L)}$, which justifies the normalization factor $\frac{m_f}{f_a}$.

Note that the ALP-fermion interactions in Eq.~\eqref{eq:ALPint} are proportional to the fermion mass $m_f$. Since the top quark mass $m_t$ is significantly larger than the masses of all other fermions in the SM, we adopt the following simplified form of the Lagrangian in subsequent analysis:
\begin{align}
    \mathcal{L}^a = &- \frac{m_t}{f_a} c_t \,a\bar{\psi}_t \psi_t - i \frac{m_t}{f_a} \tilde{c}_t\, a\bar{\psi}_t \gamma_5 \psi_t\nonumber\\
    =& -\frac{m_t}{f_a} \,a \bar{\psi}_t \big(c_t + i \gamma_5 \tilde{c}_t\big) \psi_t~.
    \label{eq:ALP-topint}
\end{align}

Similar to the Higgs-top coupling in the SM, the ALP-top coupling, as evident from Eq.~\eqref{eq:ALP-topint}, is also proportional to $m_t$. Due to the large mass of top quark, the Higgs exhibits a top-philic nature. Several studies (e.g., Refs.~\cite{Cao:2021qqt, Maltoni:2024wyh}) have explored generic CP-even or CP-odd top-philic scalars, deriving constraints on their couplings with the top quark. A phenomenology of top-philic ALPs is discussed in Refs.~\cite{Blasi:2023hvb, Tentori:2024xju}.

\section{The Chromoelectromagnetic Dipole Moment}
\label{sec:CMDMintro}
Similar to how photon-fermion interactions are described by a Lagrangian incorporating the effective electromagnetic dipole moment (EMDM), the dynamics of quark-antiquark pairs and their interactions with gluons can be characterized by a Lagrangian that includes the effective chromoelectromagnetic dipole moment (CEMDM).
The effective Lagrangian incorporating the quark CEMDM, quark EMDM, and the Weinberg operator can be expressed as~\cite{Weinberg:1989dx, Gunion:1990iv, Haberl:1995ek, Kamenik:2011dk, Bernreuther:2013aga, CMS:2016piu, DiLuzio:2020oah}:
\begin{align}
    \mathcal{L}_{\rm eff} =& -\frac{1}{2}\bar{\psi}_{q,A} \sigma^{\mu\nu} \big(\mu_q^C + i \gamma_5 d_q^C\big) \psi_{q,B} G^a_{\mu\nu} T^a_{AB}\nonumber\\
    &-\frac{1}{2} \bar{\psi}_q \sigma^{\mu\nu} \big(\mu_q^E + i\gamma_5 d^E_q\big) \psi_q F_{\mu\nu}\nonumber\\
    & -\frac{1}{6} \mathcal{W} f^{abc} \epsilon^{\mu\nu\rho\sigma} G^a_{\mu,\alpha} G^{b,\alpha}_\nu G^c_{\rho\sigma}~.
    \label{Eq:effCEMDM}
\end{align}

Here, $\sigma^{\mu\nu} = \frac{i}{2} [\gamma^\mu, \gamma^\nu]$, where $\gamma^\mu$ are the Dirac gamma matrices. In the first term, the CMDM, denoted by $\mu_q^C$, represents the CP-conserving contribution, while the CEDM, $d_q^C$, corresponds to the CP-violating part. Similarly, in the second term, the magnetic dipole moment (MDM) and EDM are denoted by $\mu_q^E$ and $d_q^E$, respectively. The CP-violating Weinberg operator is represented by the third term, where $\mathcal{W}$ denotes the Wilson coefficient. In this term, the antisymmetric Levi-Civita tensor $\epsilon^{\mu\nu\rho\sigma}$ introduces CP violation. The color generators of $\text{SU}(3)_C$ are given by $T^a_{AB}$, where $A$ and $B$ label the quark color indices, and $a$ represents the gluon color index. The field strength tensors corresponding to the photon and the gluon are defined as follows:
\begin{align}
    F_{\mu\nu} =& \partial_\mu A_\nu - \partial_\nu A_\mu~,\nonumber\\
    G_{\mu\nu}^a =& \partial_\mu g_\nu^a- \partial_\nu g_\mu^a -g_sf^{abc}g_{\mu,b} g_{\nu,c}~,
\end{align}
where $g_s=\sqrt{4\pi\alpha_s}$ is the strong coupling constant and $f^{abc}$ are the structure constants of $SU(3)_C$.
In the SM, CMDM arises at the one-loop level~\cite{Choudhury:2014lna, Bermudez:2017bpx, Martinez:2007qf, Hernandez-Juarez:2020drn}, whereas the CEDM emerges only at three-loop order~\cite{Czarnecki:1997bu}, making it significantly suppressed.
In standard conventions, the EMDM and CEMDM are typically expressed in dimensionless form as follows~\cite{ParticleDataGroup:2020ssz, Bernreuther:2013aga, CMS:2016piu, Haberl:1995ek}:
\begin{align}
    \hat{\mu}_q^E &= \left(\frac{m_q}{e}\right) \mu_q^E~,\hspace{1cm} \hat{d}_q^E = \left(\frac{m_q}{e}\right) d_q^E~,\nonumber\\
    \hat{\mu}_q^C &= \left(\frac{m_q}{g_s}\right) \mu_q^C~,\hspace{1cm} \hat{d}_q^C = \left(\frac{m_q}{g_s}\right) d_q^C~,  
    \label{eq:cemdm}
\end{align}
where $m_q$ is the mass of quark. 
The CEMDM can develop absorptive contributions, especially in regimes where the momentum transfer becomes timelike ($q^2>0$)~\cite{Hernandez-Juarez:2021xhy}. This is particularly evident in top-quark pair production ($t\bar{t}$) in $pp$ collisions~\cite{Bernreuther:2013aga, CMS:2016piu}.

Based on Eq.~\eqref{Eq:effCEMDM}, we can express the effective vertex responsible for generating the EMDM and CEMDM in the following way:
\begin{align}
    \Gamma_E^\mu &= \sigma^{\mu\nu} q_\nu \big(\mu_q^E+ i \gamma_5 d_q^E \big)~,\nonumber\\
    \Gamma_C^\mu &= \sigma^{\mu\nu} q_\nu \big(\mu_q^C+ i \gamma_5 d_q^C \big) T^a_{AB}~, 
\end{align}
where $q_\nu$ is the photon (gluon) momentum transfer for the vertex $\Gamma^\mu_{E(C)}$. One can write the corresponding invariant amplitude as follows:
\begin{align}
    \mathcal{M}_{E(C)} = \mathcal{M}^\mu_{E(C)} \epsilon^{E(C)}_\mu(\Vec{q})~,
\end{align}
where $p$ and $p^\prime$ are the momenta of the external quarks, with $q = p^\prime - p$. The Lorentz structure can be written as:
\begin{align}
    \hspace{1cm}\mathcal{M}_{E(C)}^\mu = \bar{u}(p^\prime)\Gamma^\mu_{E(C)} u(p)~.
\end{align}


\section{Calculation of the Top Quark CEDM}
\label{sec:twoloopcal}
In this section, we discuss the one-loop and two-loop computations of the top quark CEDM induced by the ALP. Note that in the following calculations, we take the external gluon to be off-shell.

\subsection{One-loop contribution}
\begin{figure}[H]
	\centering
    \includegraphics[width=0.50\linewidth]{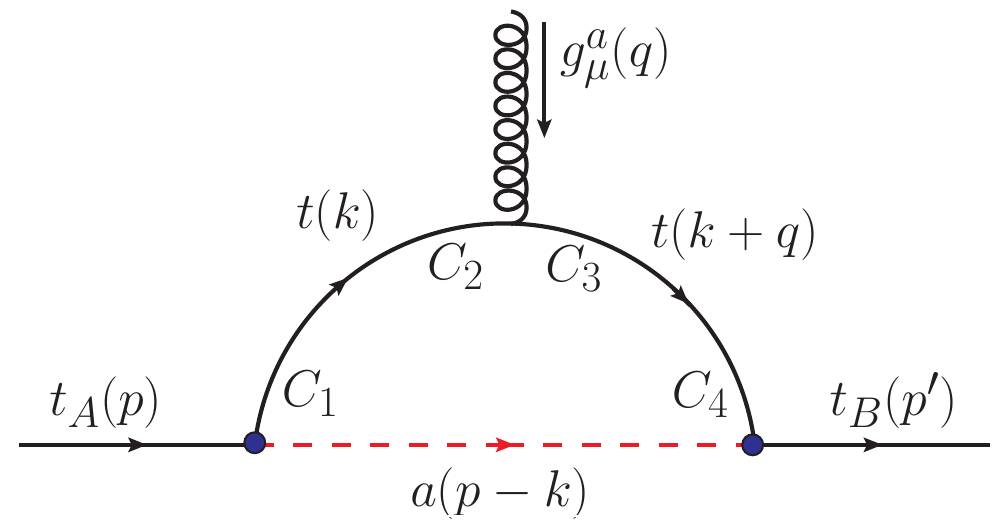}
	\caption{One-loop contributions to the top quark CEDM induced by the ALP. The ALP propagator is shown as a red line, with ALP-top quark interaction vertices indicated by blue dots.}
	\label{fig:oneloopcedm}
\end{figure}

Fig.~\ref{fig:oneloopcedm} shows the diagram contributing to the top quark CEDM at one-loop level. The ALP propagator is depicted as a red line, while ALP-top quark interaction vertices are
marked by blue dots.

The one-loop amplitude associated with the diagram in Fig.~\ref{fig:oneloopcedm} takes the form:
\begin{align}
\mathcal{M}^{\mu}_{\rm 1L}= &\int \frac{d^dk}{(2\pi)^d} \bar{u}(p^\prime)\Bigg[\bigl\{-iy_{att}(c_t + i\gamma_5 \tilde{c}_t)\delta_{C_4B}\bigr\}\Biggl\{\frac{i\big(\slashed{k}+\slashed{q}+m_{t}\big)}{\big(k+q\big)^2-m_{t}^2}\delta_{C_3C_4}\Biggr\}\big(-ig_s T^a_{C_2C_3}\gamma^\mu\big)\nonumber\\
&\times\Biggl\{\frac{i\big(\slashed{k}+m_{t}\big)}{k^2-m_{t}^2}\delta_{C_1C_2}\Biggr\}\bigl\{-iy_{att}(c_t + i\gamma_5 \tilde{c}_t)\delta_{AC_1}\bigr\}\Biggl\{\frac{i}{\big(p-k\big)^2-m_{a}^2}\Biggr\}\Bigg]u\big(p\big)~,
\label{eq:1_1La}
\end{align}
where $y_{att} = \frac{m_t}{f_a}$. 
After evaluating the loop integral, we extract the coefficient of $\gamma_5 \sigma^{\mu\nu}q_{\nu}$ to obtain the CEDM. Using Eq.~\eqref{eq:cemdm}, the top quark CEDM can then be expressed as:
\begin{align}
    \hat{d}_t^C(q^2)\Big|_{\text{1L}} =& \frac{4 y_{att}^2 m_t^2}{(4\pi)^2}c_t\tilde{c}_t\Bigg[\frac{1}{4m_t^2 - q^2}\Biggl\{f_1\big[m_t^2, m_t, m_a\big] - f_1\big[q^2, m_t, m_t\big]\nonumber\\
    &+ \frac{1}{2}\left(\frac{m_a^2}{m_t^2}\right) \log\left(\frac{m_a^2}{m_t^2}\right) - m_a^2 f_2\big[m_t^2, m_t^2, q^2, m_t, m_a, m_t\big]\Biggr\}\Bigg]~,
\end{align}
where the generic forms of the functions $f_1\big[m_1, m_2, m_3\big]$ and $f_2\big[s_1, s_2, s_3, m_1, m_2, m_3\big]$ are defined in Appendix~\ref{eq:twoloop_formfac}. We can easily evaluate the top quark CEDM at $q^2=0$ as follows:

    \begin{align}
    \hat{d}_t^C(0)\Big|_{\text{1L}} =& \frac{2 y_{att}^2}{(4\pi)^2}c_t\tilde{c}_t\Bigg[1 + \left(\frac{2 m_t^2 - m_a^2}{4m_t^2 -m_a^2}\right) f_1\big[m_t^2, m_t, m_a\big] + \frac{1}{2}\left(\frac{m_a^2}{m_t^2}\right)\log\left(\frac{m_t^2}{m_a^2}\right) \Bigg]~.
\end{align}

\subsection{Two-loop contribution}
\begin{figure}[H]
	\centering
	\includegraphics[width=0.45\linewidth]{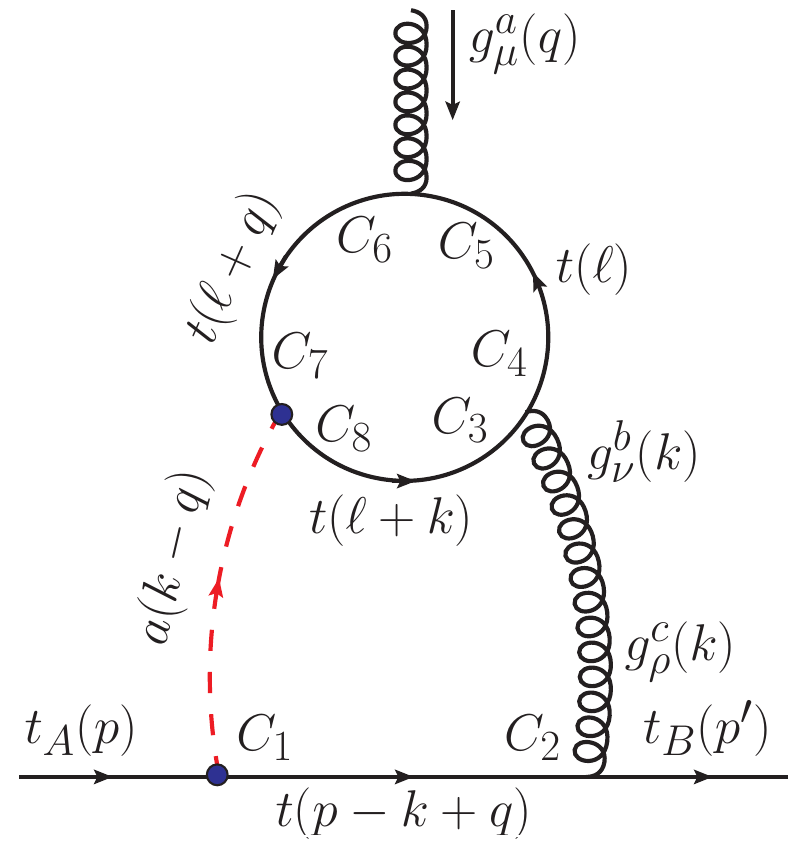}
	\caption{Two-loop contributions to the top quark CEDM induced by the ALP. The color coding is the same as in Fig.~\ref{fig:oneloopcedm}.}
	\label{fig:twoloopcedm}
\end{figure}

This section details the computation of the two-loop Barr-Zee-type contributions to the top quark CEDM. The relevant Feynman diagram, including momentum assignments, is shown in Fig.~\ref{fig:twoloopcedm}, with color coding consistent with Fig.~\ref{fig:oneloopcedm}. The calculation follows the methodology described in Refs.~\cite{Barr:1990vd, Bisal:2024nbb, Bisal:2022nbn}. 

The complete two-loop amplitude corresponding to the diagram in Fig.~\ref{fig:twoloopcedm} can be expressed as follows: 
\begin{align}
\mathcal{M}_{\rm 2L}^{\mu} =& \int\frac{d^dk}{(2\pi)^d}\int\frac{d^d\ell}{(2\pi)^d}\bar{u}(p^\prime)\Bigg[\big(-ig_sT^c_{C_2B}\gamma^\rho\big)\Biggl\{\frac{i\big(\slashed{p}-\slashed{k}+\slashed{q}+m_{t}\big)}{\big(p-k+q\big)^2-m_{t}^2}\delta_{C_1C_2}\Biggr\}\bigl\{-iy_{att}\nonumber\\&\times(c_t + i\gamma_5 \tilde{c}_t) \delta_{C_1A}\bigr\}\Biggl\{\frac{i}{\big(k-q\big)^2-m_{a}^2}\Biggr\} \Biggl\{-{\rm Tr}\Bigg[\big(-ig_sT^a_{C_5C_6}\gamma^\mu\big)\Biggl\{\frac{i\big(\slashed{\ell}+m_{t}\big)}{\ell^2-m_{t}^2}\nonumber\\
&\times\delta_{C_4C_5}\Biggr\} \big(-ig_sT^b_{C_3C_4}\gamma^\nu\big)\Biggl\{\frac{i\big(\slashed{\ell}+\slashed{k}+m_{t}\big)}{\big(\ell+k\big)^2-m_{t}^2}\delta_{C_8C_3}\Biggr\}\bigl\{-iy_{att}(c_t +i\gamma_5 \tilde{c}_t)\delta_{C_7C_8}\bigr\}\nonumber\\
&\times\Biggl\{\frac{i\big(\slashed{\ell}+\slashed{q}+m_{t}\big)}{\big(\ell+q\big)^2-m_{t}^2}\delta_{C_6C_7}\Biggr\}\Bigg]\Biggr\}\Bigg(\frac{-ig_{\nu\rho}}{k^2}\delta_{bc}\Bigg)\Bigg]u(p)~.
\label{Eq:2loopamp}
\end{align}

We can reformulate Eq.~\eqref{Eq:2loopamp} in the following form:
\begin{align}
\mathcal{M}_{\rm 2L}^{\mu}= \mathcal{K}_1\int\frac{d^dk}{(2\pi)^d}\bar{u}(p^\prime)\Bigg[\Biggl\{\frac{\gamma_\nu\big(\slashed{p}-\slashed{k}+\slashed{q}+m_{t}\big) \big(c_t + i\gamma_5 \tilde{c}_t\big)}{\mathcal{D}_1\mathcal{D}_2\mathcal{D}_3}\Biggr\}\big[\Sigma^{\mu\nu}(k)\big]\Bigg]u(p)~,
\label{Eq:7redef}
\end{align}
where $\mathcal{K}_1 = -iy_{att}g_sT^c_{C_2B}\delta_{C_1C_2}\delta_{C_1A}\delta_{bc}= -iy_{att}g_sT_{AB}^b$ and $\mathcal{D}_1= k^2$, $\mathcal{D}_2= \big(k-q\big)^2-m_{a}^2$, and $\mathcal{D}_3=\big(p-k+q\big)^2-m_{t}^2$. The quantity $\Sigma^{\mu\nu} (k)$  can be expressed as follows:

\begin{align}
\Sigma^{\mu\nu}(k) = \mathcal{K}_2\int \frac{d^d\ell}{(2\pi)^d}\Bigg[\frac{{\rm Tr}\big[\gamma^\mu\big(\slashed{\ell}+m_{t}\big)\gamma^\nu\big(\slashed{\ell}+\slashed{k}+m_{t}\big) \big(c_t + i\gamma_5 \tilde{c}_t\big) \big(\slashed{\ell}+\slashed{q}+m_{t}\big)\big]}{\mathcal{D}_4\mathcal{D}_5\mathcal{D}_6}\Bigg]~,
\label{Eq:1stloop}
\end{align}
where $\mathcal{K}_2= - y_{att}g_s^2 T^a_{C_5C_6}\delta_{C_4C_5}T^b_{C_3C_4}\delta_{C_8C_3}\delta_{C_7C_8}\delta_{C_6C_7}= -\frac{1}{2}y_{att}g_s^2\delta_{ab}$ \big(using ${\rm Tr}\big[T^aT^b\big]=\frac{1}{2}\delta^{ab}$\big) and $\mathcal{D}_4=\ell^2-m_{t}^2$, $\mathcal{D}_5=\big(\ell+k\big)^2-m_{t}^2$, and $\mathcal{D}_6=\big(\ell+q\big)^2-m_{t}^2$. 
First, we evaluate the integral $\Sigma^{\mu\nu}(k)$ over the loop momentum variable ``$\ell$''. After that, this ``$k$''-dependent result is then used to compute the second loop integral in Eq.~\eqref{Eq:7redef}.
Using Feynman parameterization, we can express Eq.~\eqref{Eq:1stloop} in the following form:

\begin{align}
\Sigma^{\mu\nu}(k) =& \mathcal{K}_2\Big[\mathcal{G}_1(k)\, g^{\mu\nu} + \mathcal{G}_2(k)\, k^\mu q^\nu + \mathcal{G}_3(k)\, k^\nu q^\mu + \mathcal{G}_4(k)\, k^{\mu} k^{\nu} \nonumber\\
&+ \mathcal{G}_5(k)\, q^\mu q^\nu + \mathcal{G}_6(k)\, \epsilon^{\mu\nu\rho\sigma} k_\rho q_\sigma \Big].
\label{Eq:9formfac}
\end{align}

The loop-momentum-dependent functions $\mathcal{G}_1(k),\,\dots,\,\mathcal{G}_6(k)$ are given by:
\begin{align}
\mathcal{G}_1(k) =& \frac{i}{\big(4\pi\big)^2}\int_{0}^{1}dx\int_{0}^{1-x}dy\,\, 4c_t m_t\left[1 - \frac{m_{t}^2}{\Delta(k,x,y)} +\frac{y^2k^2 + \big(2xy-1\big)\big(k.q\big)+ x^2 q^2}{\Delta(k,x,y)}\right]~,\nonumber\\
\mathcal{G}_2(k) =& \frac{i}{\big(4\pi\big)^2}\int_{0}^{1}dx\int_{0}^{1-x}dy\,\,4c_t m_t \left[\frac{1-4xy}{\Delta(k,x,y)}\right]~,\nonumber\\
\mathcal{G}_3(k) =& \frac{i}{\big(4\pi\big)^2}\int_{0}^{1}dx\int_{0}^{1-x}dy\,\,4c_t m_t\left[-\frac{\big(1-2x\big)\big(1-2y\big)}{\Delta(k,x,y)}\right]~,\nonumber\\
\mathcal{G}_4(k) =& \frac{i}{\big(4\pi\big)^2}\int_{0}^{1}dx\int_{0}^{1-x}dy\,\,8c_t m_t\left[\frac{y\big(1-2y\big)}{\Delta(k,x,y)}\right]~,\nonumber\\
\mathcal{G}_5(k) =& \frac{i}{\big(4\pi\big)^2}\int_{0}^{1}dx\int_{0}^{1-x}dy\,\,8c_t m_t\left[\frac{x\big(1-2x\big)}{\Delta(k,x,y)}\right]~,\nonumber\\
\mathcal{G}_6(k) =& \frac{i}{\big(4\pi\big)^2}\int_{0}^{1}dx\int_{0}^{1-x}dy\,\,4\tilde{c}_t m_t\left[\frac{1}{\Delta(k,x,y)}\right]~,
\label{eq:formfac}
\end{align}
where $\Delta(k,x,y)= x(x-1)q^2 + y(y-1)k^2 + 2xy(k.q)+ m_t^2$, $x$ and $y$ represent the Feynman parameters. By substituting the expressions for $\mathcal{G}_1(k),\, \dots,\, \mathcal{G}_6(k)$ from Eq.~\eqref{eq:formfac}, we evaluate the loop momentum integral in Eq.~\eqref{Eq:7redef} over the variable ``$k$''. This yields six different integrals, each associated with one of the functions defined in Eq.~\eqref{eq:formfac}. We then isolate the coefficient of $\gamma_5 \sigma^{\mu\nu} q_\nu$ to extract the two-loop form factors contributing to the CEDM of the top quark. Note that the terms involving $\mathcal{G}_3(k)$ and $\mathcal{G}_5(k)$ do not contribute to this coefficient, as it vanishes in both cases. Moreover, the form factor associated with $\mathcal{G}_4(k)$ also evaluates to zero. Thus, the only non-zero two-loop form factors are $\mathcal{F}_1(q^2)$, $\mathcal{F}_2(q^2)$, and $\mathcal{F}_6(q^2)$. Consequently, the CEDM of the top quark for $q^2\neq 0$ can be expressed as follows:
\begin{align}
    \hat{d}_t^C(q^2)\Big|_{\rm 2L} &= \left(\frac{m_t}{g_s}\right) d_t^C(q^2)\Big|_{\rm 2L} \nonumber\\
    &=\left(\frac{m_t}{g_s}\right) \left[\mathcal{F}_1(q)^2 + \mathcal{F}_2(q)^2 + \mathcal{F}_6(q)^2\right]~,
    \label{eq:twoloopformfac}
\end{align}
where the complete analytical expressions for the form factors $\mathcal{F}_1(q^2)$, $\mathcal{F}_2(q^2)$, and $\mathcal{F}_6(q^2)$ are provided in Appendix~\ref{eq:twoloop_formfac}.

Finally, the total CEDM (one-loop $+$ two-loop) of the top quark is given by:
\begin{align}
    \hat{d}_t^C(q^2) = \hat{d}_t^C(q^2)\Big|_{\rm 1L} +  \hat{d}_t^C(q^2)\Big|_{\rm 2L}~.
    \label{eq:totalcedm}
\end{align}

\section{Contribution to the Weinberg Operator, Neutron EDM, and Mercury EDM}
\label{sec:edmcedmweinberg}
The QCD renormalization group (RG) evolution causes the operators in Eq.~\eqref{Eq:effCEMDM} to run and mix. These effects have been calculated up to next-to-leading logarithmic (NLL) precision~\cite{Degrassi:2005zd}. Specifically, the Weinberg operator induces mixing into quark EDMs and CEDMs but the reverse does not occur. However, it has long been established that quark CEDMs generate a finite threshold correction to the Weinberg operator upon integrating out a heavy quark (top)~\cite{Braaten:1990gq, Chang:1991ry}. 

\begin{figure}[H]
	\centering
    \includegraphics[width=0.40\linewidth]{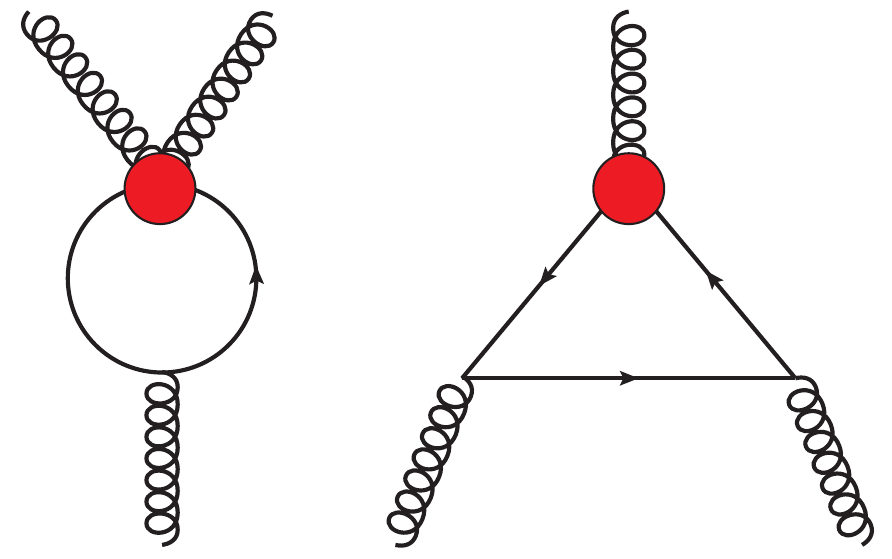}\\
    \hspace{-0.50cm}(a)\hspace{2.65cm}(b)
	\caption{Figures (a) and (b) show the diagrams contributing to the Weinberg operator at the top quark threshold, with the red blob representing the CEDM insertion.}
	\label{fig:weinberg}
\end{figure}

Figs.~\ref{fig:weinberg}a and \ref{fig:weinberg}b depict the diagrams contributing to the Weinberg operator $\mathcal{W}$ at the top quark threshold, where the red blob denotes the insertion of the CEDM of the top quark. 

One can write the threshold correction to $\mathcal{W}$ as~\cite{Kamenik:2011dk}:
\begin{align}
    \delta \mathcal{W}^{(t)} = \frac{g_s^2}{32\pi^2 m_t} d^C_t(m_t)~,
\end{align}
where $d_t^C(m_t)$ can be found from Eq.~\eqref{eq:twoloopformfac}. The finite correction in $\mathcal{W}$, along with the subsequent RG running down to hadronic energies, results in non-zero contributions to both the EDMs and CEDMs of light quarks. Following Ref.~\cite{Kamenik:2011dk}, which performs the RG running of the Weinberg operator at NLL accuracy down to the hadronic scale, $\mu_{\rm Had}\sim 1$~GeV, we can write $d_{u,d}^E(\mu_{\rm Had})$, $d^C_{u,d}(\mu_{\rm Had})$, and $\mathcal{W}(\mu_{\rm Had})$ in terms of $d^C_t(m_t)$ as follows:

\begin{align}
    d_u^E =& -3.1\times 10^{-9} e \left(\frac{d_t^C(m_t)}{g_s}\right)~, \nonumber\\
    d_d^E =& 3.5\times 10^{-9} e \left(\frac{d_t^C(m_t)}{g_s}\right)~, \nonumber\\
    d_u^C =& 8.9\times 10^{-9} d_t^C(m_t)~, \nonumber\\
    d_d^C =& 2.0 \times 10^{-8} d_t^C(m_t)~, \nonumber\\
    \mathcal{W} =& 1.0\times 10^{-5}~\text{GeV}^{-1} \left(\frac{d_t^C(m_t)}{g_s}\right)~,
    \label{eq:edmneutron}
\end{align}
where $m_t=173.3$~GeV, $m_d^{\overline{\rm MS}}(2~\text{GeV}) = (4.7\pm 0.1)$~MeV, $m_u^{\overline{\rm MS}}(2~\text{GeV}) = (2.1\pm 0.1)$~MeV, and $\alpha_s^{\overline{\rm MS}}(m_Z)=0.118$ have been used. Note that using Eq.~\eqref{eq:totalcedm}, we obtain an exact calculation of $d_t^C(m_t)$ by fixing the momentum transfer at $q^2=m_t^2$. Finally, using Eq.~\eqref{eq:edmneutron}, we can compute the EDM of neutron and mercury (Hg). The present experimental constraints on the EDM of neutron and Hg are:
\begin{align}
    \left\lvert d^E_n \right\rvert_{\text{Exp.}} &< 1.8 \times 10^{-26}e\;\text{cm}\;(90\%\;\text{C.L.})~\text{\cite{Abel:2020pzs}},\nonumber\\
    \left\lvert d_{\rm Hg}^E\right\rvert_{\text{Exp.}} &< 7.4\times 10^{-30} e\;\text{cm}\;(95\%\;\text{C.L.})~\text{\cite{Graner:2016ses}}.
    \label{eq:explimitedm}
\end{align}

Following the approach in Ref.~\cite{Pospelov:2005pr}, we may write the contributions to the neutron and Hg EDMs as follows~\cite{Pospelov:2000bw, Hisano:2012cc, Kamenik:2011dk, Cirigliano:2019vfc, Yamanaka:2018uud, Osamura:2022rak, Yamanaka:2017mef}:
\begin{align}
    d_n^E =& \big(1\pm 0.5\big) \left[1.1e\left(\frac{d_d^C}{g_s} + 0.5 \frac{d_u^C}{g_s}\right)\right] + \left(0.8\,d_d^E -0.2\, d_u^E\right)  + \big(22\pm 10\big)\times 10^{-3}~\text{GeV}\;e\,\mathcal{W}~,\nonumber\\
     d_{\rm Hg}^E =& \big(-1.3\pm 0.96\big)\times 10^{-5}~\text{GeV}\,e\,\mathcal{W}~.
    \label{eq:neutronedm}
\end{align}
Here, all quantities are calculated at the energy scale of $\mu_{\rm Had}\sim 1$~GeV. Note that the numerical values and associated uncertainties for the relevant matrix elements, particularly the Weinberg operator's contribution to the $d_n^E$, have been computed using QCD sum rule~\cite{Demir:2002gg}~\footnote{The recent calculation of the Weinberg operator's contribution to the neutron EDM can be found in Ref.~\cite{Yamanaka:2020kjo}, which agrees well with the results of Ref.~\cite{Demir:2002gg}.}. Additionally, the light quark EDM contributions to the $d_n^E$ and the Weinberg operator's contribution to the $d_{\rm Hg}^E$ have been taken from recent lattice results~\cite{Yamanaka:2018uud, Osamura:2022rak, Yamanaka:2017mef}.

\section{Constraints on ALP-Top Quark Coupling from Other Searches}
\label{sec:limitothersearches}
The study in Ref.~\cite{Esser:2023fdo} explored the coupling of a light ALP to top quarks, considering both direct and indirect probes. The direct search focused on $t\bar{t}$ production associated with an ALP, while the indirect approach investigated ALP production via the gluon fusion process, followed by decays into top pairs. The limit at $95\%$ C.L. obtained from the ATLAS search is given by
\begin{align}
    \left|\frac{c_t}{f_a}\right| < 1.81\times 10^{-3}~\text{GeV}^{-1}~.
\end{align}

The production of di-boson, mediated by an ALP through its coupling with the top quark leads to the following constraint
\begin{align}
    \left|\frac{c_t}{f_a}\right| < 4.44\times 10^{-2}~\text{GeV}^{-1}~.
\end{align}

The study of $t\bar{t}$ production involving a high-$p_T$ top quark, as measured by ATLAS, was used in Ref.~\cite{Esser:2023fdo} to derive the constraint
\begin{align}
    \left|\frac{c_t}{f_a}\right| < 5.90\times 10^{-3}~\text{GeV}^{-1}~.
\end{align}

Additional phenomenological studies probing the coupling of low-mass ALPs to the top quark can be found in Refs.~\cite{Hosseini:2024kuh, Ebadi:2019gij}.

In Ref.~\cite{Carra:2021ycg}, the study builds upon the work of Ref.~\cite{Gavela:2019cmq} by calculating the limits on the ALP-$WW$ and ALP-$Z\gamma$ couplings using ATLAS non-resonant searches in the $WW$ and $Z\gamma$ final states. Ref~\cite{Esser:2023fdo} studied the $Z\gamma$ final state and derived the following $95\%$ C.L limit (for $m_a<100$~GeV):
\begin{align}
    \left|\frac{c_t}{f_a}\right| < 9.09\times 10^{-2}~\text{GeV}^{-1}~.
\end{align}

Constraints on the ALP-$ZZ$ coupling are further studied in Ref.~\cite{CMS:2021xor} through a CMS non-resonant searches of $ZZ$ production. This leads to the constraint on ALP-top coupling at $95\%$ C.L. as follows:
\begin{align}
    \left|\frac{c_t}{f_a}\right| < 5.88\times 10^{-2}~\text{GeV}^{-1}~.
\end{align}

Based on the lowe-energy precision measurements of rare Kaon~\cite{NA62:2021zjw} and $B$-meson decays~\cite{BaBar:2013npw} from the  NA62 and BABAR collaboration, Ref.~\cite{Esser:2023fdo} derived the following limits:
\begin{align}
    \left|\frac{c_t}{f_a}\right| &< 2.80\times 10^{-4}~\text{GeV}^{-1} \; \; \text{\big($K$ decays: $m_a<110$~MeV and $m_a\in [160, 260]$~MeV\big)}~,\\
    \left|\frac{c_t}{f_a}\right| &< 1.15\times 10^{-6}~\text{GeV}^{-1} \; \; \text{\big($B$ decays: $m_a\lesssim 5$~GeV\big)}~.
\end{align}

Ref.~\cite{Phan:2023dqw} provides a comprehensive study of ALPs in top-pair production at the LHC, deriving constraints on the ALP-top coupling from the total cross section and differential distributions. The resulting bound is given by:
\begin{align}
    \left|\frac{c_t}{f_a}\right| < 7\times 10^{-3}~\text{GeV}^{-1}~, \; \; \text{\big(for $0<m_a<200$~GeV\big)}.
\end{align}

Based on the measurements of top-antitop production accompanied by di-muon resonances from the ATLAS~\cite{ATLAS:2023ofo} and CMS~\cite{CMS:2019lwf} experiments, the following constraint is derived:
\begin{align}
    \left|\frac{c_t}{f_a}\right| < 1\times 10^{-3}~\text{GeV}^{-1}~, \; \; \text{\big(LHC $150~\text{fb}^{-1}$\big)}.
\end{align}

An analysis of top quark pair production at the LHC yields the bound~\cite{Bruggisser:2023npd}: 
\begin{align}
    \left|\frac{c_t}{f_a}\right| < 4\times 10^{-3}~\text{GeV}^{-1}~, \; \; \text{\big(for $m_a=300$~GeV\big)}.
\end{align}

A related study in Ref.~\cite{Blasi:2023hvb} finds a slightly weaker limit:
\begin{align}
    \left|\frac{c_t}{f_a}\right| < 6\times 10^{-3}~\text{GeV}^{-1}~, \; \; \text{\big($10~\text{GeV}<m_a<200$~GeV\big)}.
\end{align}

\section{Constraints on ALP-Top Coupling from Top Quark CEDM, Neutron EDM, and Mercury EDM}
\label{sec:mainresults}
In Sec.~\ref{sec:twoloopcal} and \ref{sec:edmcedmweinberg}, we have already discussed the relevant formulae for computing the CEDM and EDMs. In this section, we present the numerical values of the constraints on the ALP-top coupling derived from the top quark CEDM, the neutron EDM, and the mercury EDM. The top quark CEDM, evaluated up to two-loop order at $q^2=m_t^2$, is subsequently used to compute the corresponding contributions to the EDMs of the neutron and mercury. We provide the resulting bounds on the ALP-top coupling for various values of the ALP mass, $m_a$, as obtained from each observable.

In our scenario, it is important to note that the CEDM of the top quark, and consequently the EDM of the neutron and mercury, develops an absorptive part for $q^2 > 0$. Specifically, it contains both real and imaginary components.

\begin{figure}[H]
	\centering
    \includegraphics[width=0.65\linewidth]{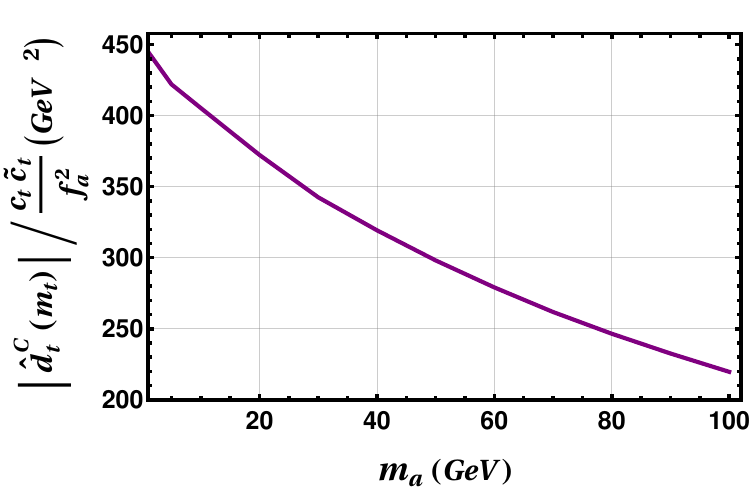}
	\caption{Variation of the CEDM of the top quark in unit of $\frac{c_t\tilde{c}_t}{f_a^2}$ with the ALP mass $m_a$.}
	\label{fig:cedmplot}
\end{figure}

\vspace{0.3cm}
\noindent
\textbf{Constraints from CEDM of top quark:} 
In Fig.~\ref{fig:cedmplot}, we show the variation of $\hat{d}_t^C(m_t)/\left(\frac{c_t \tilde{c}_t}{f_a^2}\right)$ with the ALP mass, which is varied over the range $m_a \in [1,100]$~GeV. As an example, for $m_a=1$~GeV, we have:
\begin{align}
    \left\lvert \hat{d}_t^C(m_t)\right\rvert = 445\times \left(\frac{c_t \tilde{c}_t}{f_a^2}\right)~.
    \label{eq:valuecmdm50}
\end{align}

Applying the latest experimental bound on the top quark CEDM given in Eq.~\eqref{eq:cedmbound}, we obtain from Eq.~\eqref{eq:valuecmdm50}:
\begin{align}
    \frac{c_t}{f_a}\cdot \frac{\tilde{c}_t}{f_a} <6.74\times 10^{-5}~\text{GeV}^{-2}~.
\end{align}

In the case of maximal CP violation, where $c_t\simeq \tilde{c}_t$, we obtain:
\begin{align}
    \left\lvert \frac{c_t}{f_a}\right\rvert< 8.21\times 10^{-3}~\text{GeV}^{-1} \implies \left\lvert\frac{f_a}{c_t}\right\rvert> 122~\text{GeV}~.
\end{align}

Similarly, we can derive the limit for $m_a=100$~GeV as:
\begin{align}
   \left\lvert \hat{d}_t^C(m_t)\right\rvert = 220\times \left(\frac{c_t \tilde{c}_t}{f_a^2}\right)~.
\end{align}

Therefore, we obtain:
\begin{align}
    \left\lvert \frac{c_t}{f_a}\right\rvert< 1.17\times 10^{-2}~\text{GeV}^{-1} \implies \left\lvert\frac{f_a}{c_t}\right\rvert> 86~\text{GeV}~.
\end{align}

Thus, the constraint derived from the top quark CEDM is found to be more stringent than those from di-boson production, ATLAS non-resonant searches in the $Z\gamma$ final state, and CMS non-resonant $ZZ$ production searches. In fact, the limit from top quark CEDM is nearly an order of magnitude stronger than the bounds set by the ATLAS $Z\gamma$ and CMS $ZZ$ non-resonant searches.


\vspace{0.3cm}
\noindent
\textbf{Constraints from EDM of neutron:} 
Using Eq.~\eqref{eq:neutronedm}, we calculate the neutron EDM for an ALP mass of $m_a=1$~GeV as:

\begin{align}
    \left|d_n^E\right| = 7\times 10^{-21} \times \left(\frac{c_t \tilde{c}_t}{f_a^2}\right) e~\text{cm}~.
\end{align}

Using the experimental limit on $\left|d_n^E\right|$ given in Eq.~\eqref{eq:explimitedm}, we obtain:
\begin{align}
  \frac{c_t}{f_a}\cdot \frac{\tilde{c}_t}{f_a} < 2.57\times 10^{-6}~\text{GeV}^{-2}~.  
\end{align}

With $c_t\simeq \tilde{c}_t$, this gives:
\begin{align}
    \left|\frac{c_t}{f_a}\right| < 1.6\times 10^{-3}~\text{GeV}^{-1} \implies \left|\frac{f_a}{c_t}\right|> 625~\text{GeV}~.
\end{align}

In the same way, the limit for $m_a=100$~GeV can be obtained as follows:
\begin{align}
    \left|d_n^E\right| = 3.45\times 10^{-21} \times \left(\frac{c_t \tilde{c}_t}{f_a^2}\right) e~\text{cm}~.
\end{align}

Therefore,
\begin{align}
    \left|\frac{c_t}{f_a}\right| < 2.28\times 10^{-3}~\text{GeV}^{-1} \implies \left|\frac{f_a}{c_t}\right|> 439~\text{GeV}~.
\end{align}


Thus, the constraint on the ALP-top coupling derived from the neutron EDM is more than five times stronger than that from the top quark CEDM. Moreover, it surpasses bounds derived from ATLAS searches for $t\bar{t}$ production associated with an ALP, high-$p_T$ top quark measurements, and analyses of total cross-sections and differential distributions in $t\bar{t}$ production at the LHC. 

\vspace{0.2cm}
\noindent
\textbf{Constraints from EDM of mercury:}
We employ Eq.~\eqref{eq:neutronedm} to evaluate the mercury EDM. For $m_a = 1$~GeV, we obtain:

\begin{align}
    \left|d_{\rm Hg}^E\right| = 1.73\times 10^{-24} \times \left(\frac{c_t \tilde{c}_t}{f_a^2}\right) e~\text{cm}~.
\end{align}

From the experimental constraint on $\left|d_{\rm Hg}^E\right|$ as noted in Eq.~\eqref{eq:explimitedm}, we obtain:
\begin{align}
  \frac{c_t}{f_a}\cdot \frac{\tilde{c}_t}{f_a} < 4.27\times 10^{-6}~\text{GeV}^{-2}~.  
\end{align}

For $c_t\simeq \tilde{c}_t$, this yields:
\begin{align}
    \left|\frac{c_t}{f_a}\right| < 2.06\times 10^{-3}~\text{GeV}^{-1} \implies \left|\frac{f_a}{c_t}\right|> 485~\text{GeV}~.
\end{align}

Similarly, for $m_a=100$~GeV, we obtain:
\begin{align}
    \left|d_{\rm Hg}^E\right| = 8.56\times 10^{-25} \times \left(\frac{c_t \tilde{c}_t}{f_a^2}\right) e~\text{cm}~.
\end{align}

Therefore, we obtain the following limit:
\begin{align}
    \left|\frac{c_t}{f_a}\right| < 2.94\times 10^{-3}~\text{GeV}^{-1} \implies \left|\frac{f_a}{c_t}\right|> 340~\text{GeV}~.
\end{align}

Thus, the constraint on the ALP-top coupling obtained from the mercury EDM is nearly four times stronger than that from the top quark CEDM but remains weaker than the bound derived from the neutron EDM.

\section{Summary and Conclusions}
In this work, we consider an ALP that arises as a mixture of scalar and pseudoscalar operators. Due to this mixed nature, the ALP induces CP violation when it couples to the SM fermions. Such CP violation can generate EDMs for quarks and leptons and CEDMs for quarks. We compute the CEDM of the top quark mediated by this ALP at one-loop and two-loop levels, where the two-loop contribution stems from the Barr-Zee-type diagrams. A key distinction is that, unlike the static EDMs of fermions, which are constrained by low-energy precision experiments, the CEDM can only be probed dynamically in high-energy collisions. To retain generality, we consider the external gluon to be off-shell, i.e., $q^2 \neq 0$, and ultimately evaluate the CEDM at $q^2 = m_t^2$, corresponding to the pole mass of the top quark.
Since the top quark CEDM can induce finite threshold corrections to the Weinberg operator upon integrating out the top quark, we have computed its contribution to the Weinberg operator explicitly. This finite correction, combined with the subsequent RG running down to the hadronic scale, generates non-zero contributions to both the EDMs and CEDMs of light quarks. We then express the EDMs and CEDMs of light quarks at the hadronic scale in terms of the top quark CEDM evaluated at its pole mass. These results are subsequently used to calculate the EDMs of the neutron and mercury. Using the current experimental upper limits on the top quark CEDM and the EDMs of the neutron and mercury, we derive bounds on the ALP-top coupling. 
Our analysis shows that the constraint derived from the neutron EDM is significantly more stringent than those from the top quark CEDM and the mercury EDM.
From the neutron EDM constraint, we obtain a bound on the ALP-top coupling: $\left|\frac{c_t}{f_a}\right| < 1.6\times 10^{-3}~\text{GeV}^{-1}$ for an ALP mass of  $m_a=1$~GeV. 
Thus, our bound on the ALP-top coupling, \( \left|\frac{c_t}{f_a}\right| \), is stronger than the limits derived from the ATLAS \( t\bar{t}a \) search and the ATLAS measurement of high-\( p_T \) \( t\bar{t} \) production.

\label{sec:conclusion}

\section{Acknowledgements}
I would like to thank U. Chattopadhyay for carefully reading the manuscript and for the valuable discussions. I also thank D. Das for helpful discussions.

\appendix
\section*{Appendix}
\section{Two-Loop Form Factors}
\label{eq:twoloop_formfac}
The two-loop form factors contributing to the CEDM of the top quark, as used in Eq.~\eqref{eq:twoloopformfac}, can be expressed as follows:
\begin{align}
\mathcal{F}_1(q^2)=&-\frac{2y_{att}^2 g_s^3 m_t}{(4\pi)^4} c_t \tilde{c}_t\int_{0}^1 dx\int_{0}^{1-x}dy\,\Bigg[\frac{8 m_t^2 f_1[m_t^2, m_t, m_a]}{q^2(q^2-4 m_t^2 )} + \frac{4 (m_a^2 - q^2) }{q^2(q^2-4 m_t^2 )}  \log\left(\frac{m_t^2}{m_a^2 - q^2}\right) \nonumber\\
& -\frac{4 \left\{-m_a^2 q^2 + 2 m_t^2 (m_a^2 + q^2)\right\} f_2\left[m_t^2, m_t^2, q^2, m_a, m_t, 0\right]}{q^2(q^2-4 m_t^2 )} - \frac{2m_t^2}{M^2+ \xi\left\{m_a^2- q^2(1+\xi)\right\}}\nonumber\\
& \times \Biggl\{-\frac{2 q^2 (1 + \xi)^3 f_1\left[q^2 (1 + \xi)^2, M, m_a\right]}{\Lambda\left[m_t^2, q^2 (1 + \xi)^2, m_t^2 + q^2 \xi (1 + \xi)\right]} 
- \frac{\left(-M^2 + m_a^2 + q^2 (-1 + \xi^2)\right) \log\left(\frac{M^2}{m_a^2}\right)}{q^2 (q^2-4 m_t^2) (1 + \xi)} \nonumber\\
& +\frac{2 \left(-m_a^2 + q^2\right) \log\left(\frac{m_a^2}{m_a^2 - q^2}\right)}{q^2(q^2-4 m_t^2 )} + \frac{2 \xi \left(-M^2 + q^2 \xi^2\right) \log\left(\frac{M^2}{M^2 - q^2 \xi^2}\right)}{\Lambda\left[m_t^2, q^2 \xi^2, m_t^2 + q^2 \xi (1 + \xi)\right]} \nonumber\\
&+\frac{2 \left\{-m_a^2 q^2 + 2 m_t^2 \left(m_a^2 + q^2\right)\right\} f_2\left[m_t^2, m_t^2, q^2, m_a, m_t, 0\right]}{q^2(q^2-4 m_t^2 )} \nonumber\\
& +\frac{2 \xi^2 \left\{q^2 \left(M^2 - q^2 \xi (1 + \xi)\right) + m_t^2 \left(-2 M^2 + 2 q^2 \xi (2 + \xi)\right)\right\}}{\Lambda\left[m_t^2, q^2 \xi^2, m_t^2 + q^2 \xi (1 + \xi)\right]}\nonumber\\
& \times f_2\left[m_t^2, q^2 \xi^2, m_t^2 + q^2 \xi (1 + \xi), m_t, 0, M\right] \nonumber\\
& - \frac{2 (1 + \xi)^2 \left\{-m_a^2 q^2 (1 + \xi) + 2 m_t^2 \left(-M^2 + m_a^2 + q^2 (1 + \xi)^2\right)\right\} }{\Lambda\left[m_t^2, q^2 (1 + \xi)^2, m_t^2 + q^2 \xi (1 + \xi)\right]}\nonumber\\
& \times f_2\left[m_t^2, q^2 (1 + \xi)^2, m_t^2 + q^2 \xi (1 + \xi), m_t, m_a, M\right]\Biggr\}\left\{\frac{1}{y(y-1)}\right\} \nonumber\\
& + \left\{\frac{1}{y(y-1)}\right\}\Biggl\{\frac{4 m_t^2 \left(-1 + 2 y^2 + 2 y z\right) (1 + \xi) f_1\left[m_t^2, m_t, m_a\right]}{\Lambda\left[m_t^2, q^2 (1 + \xi)^2, m_t^2 + q^2 \xi (1 + \xi)\right]} \nonumber\\
& +\frac{2 q^2 (1 + \xi)^2 f_1\left[q^2 (1 + \xi)^2, M, m_a\right]}{\left(M^2 + \xi (m_a^2 - q^2 (1 + \xi))\right) \Lambda\left[m_t^2, q^2 (1 + \xi)^2, m_t^2 + q^2 \xi (1 + \xi)\right]}\nonumber\\
&\times \Bigl\{M^2 (1 - 2 y^2 - 2 y z) + m_a^2 (-1 + 2 y z - 2 y^2 \xi) + q^2 (1 + \xi) (1 - 2 z^2 + 2 y z (-1 + \xi) \nonumber\\
&- \xi + 2 y^2 \xi)\Bigr\} +\frac{2 q^2 \left(q^2-4 m_t^2\right) \xi (1 + \xi) \left(1 - 2 x y + 2 y^2 \xi\right) }{\Lambda\left[m_t^2, q^2 \xi^2, m_t^2 + q^2 \xi (1 + \xi)\right] \Lambda\left[m_t^2, q^2 (1 + \xi)^2, m_t^2 + q^2 \xi (1 + \xi)\right]}\nonumber\\
& \times \left\{-2 m_t^2 + q^2 (1 + \xi)\right\} f_1\left[m_t^2 + q^2 \xi (1 + \xi), m_t, M\right] \nonumber\\
&- \frac{2 m_t^2 \left\{m_a^2 (1 - 2 y^2 - 2 x y ) \xi + M^2 (1 - 2 x y  + 2 y^2 \xi) + q^2 \xi (1 + \xi) (1 - 2 x y + 2 y^2 (1 + \xi))\right\} }{q^2 (q^2-4 m_t^2) \xi (1 + \xi) (m_t^2 + q^2 \xi (1 + \xi))}\nonumber\\
&\times \log\left(\frac{m_t^2}{M^2}\right) + \frac{ 1}{(q^2 -4 m_t^2) \xi (m_t^2 + q^2 \xi (1 + \xi))}\log\left(\frac{m_t^2}{M^2}\right)\nonumber\\
&\times \left\{M^2 (1 - 2 x y + 2 y^2 \xi) + \xi \left(2 m_a^2 (-1 + 2 y^2 + 2 x y) \xi - q^2 (1 + \xi) (-1 + 2 x y + 2 y^2 \xi)\right)\right\} \nonumber\\
& + \frac{1}{q^2 (q^2 - 4 m_t^2) (1 + \xi)^2 (M^2 + \xi (m_a^2 - q^2 (1 + \xi)))} \Big\{M^4 (-1 + 2 y^2 + 2 x y) \nonumber\\
&+ q^4 (1 + \xi)^2 (-1 - 2 x^2 (-1 + \xi) - \xi^2 + 2 y^2 \xi (1 + \xi) + 2 x y (1 + \xi^2)) - 2 m_a^2 q^2 (1 + \xi) \nonumber\\
&\times (-1 + x^2 - \xi - \xi^2 + y^2 \xi (2 + 3 \xi) + 2 x y (1 + \xi + \xi^2)) + m_a^4 (-1 - 2 \xi - 2 \xi^2 + 2 y^2 \xi (1 + 2 \xi) \nonumber\\
&+ 2 x y (1 + 2 \xi + 2 \xi^2)) + 2 M^2 (-q^2 (1 + \xi) (-x^2 - \xi + 2 x y \xi + y^2 (1 + 2 \xi)) \nonumber\\
&+ m_a^2 (-\xi + 2 x y \xi + y^2 (1 + 3 \xi))))\Big\} \log\left(\frac{M^2}{m_a^2}\right) \nonumber\\
& +\frac{2 \left(-m_a^2 + q^2\right) \left\{m_a^2 (1 - 2 x y) + q^2 (-1 + 2 x y + 2 x^2)\right\} }{q^2 \left(-4 m_t^2 + q^2\right) \left(M^2 + \xi \left(m_a^2 - q^2 (1 + \xi)\right)\right)}\log\left(\frac{m_a^2}{m_a^2 - q^2}\right) \nonumber\\
& - \frac{2 \left(M^2 - q^2 \xi^2\right) \left\{M^2 (2 x y - 1) + q^2 \xi \left(2 x^2 + \xi - 2 x y \xi\right)\right\} }{\left(M^2 + \xi \left(m_a^2 - q^2 (1 + \xi)\right)\right) \Lambda[m_t^2, q^2 \xi^2, m_t^2 + q^2 \xi (1 + \xi)]} \log\left(\frac{M^2}{M^2 - q^2 \xi^2}\right)  \nonumber\\
& - \frac{2 \left\{2 m_t^2 \left(m_a^2 + q^2\right) - m_a^2 q^2\right\} \left\{m_a^2 ( 2 x y - 1) + q^2 \left(1 - 2 x y - 2 x^2\right)\right\} f_2\left[m_t^2, m_t^2, q^2, m_a, m_t, 0\right]}{q^2 \left(q^2 -4 m_t^2 \right) \left\{M^2 + \xi \left(m_a^2 - q^2 (1 + \xi)\right)\right\}} \nonumber\\
& + \frac{2 \xi \left\{M^2 (1 - 2 x y) + q^2 \xi \left(-2 x^2 - \xi + 2 x y \xi\right)\right\}}{\left\{M^2 + \xi \left(m_a^2 - q^2 (1 + \xi)\right)\right\} \Lambda\left[m_t^2, q^2 \xi^2, m_t^2 + q^2 \xi (1 + \xi)\right]} \nonumber\\
& \times \left\{q^4 \xi (1 + \xi) - M^2 q^2 + 
2 m_t^2 (M^2 - q^2 \xi (2 + \xi))\right\} f_2[m_t^2, q^2 \xi^2, 
m_t^2 + q^2 \xi (1 + \xi), m_t, 0, M] \nonumber\\
&+ \frac{ M^2 (1 - 2 y^2 - 2 x y) + m_a^2 (2 x y - 2 y^2 \xi-1) + q^2 (1 + \xi) \left\{1 - 2 x^2 + 2 x y (\xi-1) - \xi + 2 y^2 \xi\right\}}{\left\{M^2 + \xi (m_a^2 - q^2 (1 + \xi))\right\} \Lambda\left[m_t^2, q^2 (1 + \xi)^2, m_t^2 + q^2 \xi (1 + \xi)\right]}\nonumber\\
& \times 2 (1 + \xi) \left\{2 m_t^2 (m_a^2 - M^2 + q^2 (1 + \xi)^2) -m_a^2 q^2 (1 + \xi)\right\} \nonumber\\
&\times f_2[m_t^2, q^2 (1 + \xi)^2, m_t^2 + q^2 \xi (1 + \xi), m_t, m_a, M]\Biggr\} \Bigg]~,
\end{align}

\begin{align}
\mathcal{F}_2(q^2) =& -\frac{2y_{att}^2 g_s^3 m_t}{(4\pi)^4} c_t \tilde{c}_t\int_{0}^1 dx\int_{0}^{1-x}dy\,\left[\frac{1-4xy}{y(y-1)}\right]\,\nonumber\\
&\times\Bigg[\frac{4 m_a^2 q^4 (q^2-4 m_t^2) (1 + \xi)^5 f_1\left[q^2 (1 + \xi)^2, M, m_a\right]}{\left\{M^2 + \xi (m_a^2 - q^2 (1 + \xi))\right\} \Lambda[m_t^2, q^2 (1 + \xi)^2, m_t^2 + q^2 \xi (1 + \xi)]^2} \nonumber\\
& - \frac{2 q^8 \left(q^2-4 m_t^2\right)^3 \xi^3 (1 + \xi)^4 (1 + 2 \xi) f_1\left[m_t^2 + q^2 \xi (1 + \xi), m_t, M\right]}{\Lambda\left[m_t^2, q^2 \xi^2, m_t^2 + q^2 \xi (1 + \xi)\right]^2 \Lambda\left[m_t^2, q^2 (1 + \xi)^2, m_t^2 + q^2 \xi (1 + \xi)\right]^2}\nonumber\\
& + \frac{-4 m_t^2 \xi + M^2 (1 + 2 \xi) - q^2 \xi (-1 + \xi + 2 \xi^2) }{(4 m_t^2 - q^2) \xi (m_t^2 + q^2 \xi (1 + \xi))}\log\left(\frac{m_t^2}{m_a^2}\right)\nonumber\\
& + \frac{M^4 q^2 (1 + 3 \xi + 2 \xi^2) - M^2 \xi \left\{2 m_t^2 (m_a^2 + 2 q^2 (1 + \xi)) + q^2 (1 + \xi) (4 q^2 \xi (1 + \xi)-m_a^2)\right\} }{q^2 (q^2-4 m_t^2) \xi (1 + \xi) (m_t^2 + q^2 \xi (1 + \xi)) \left\{M^2 + \xi (m_a^2 - q^2 (1 + \xi))\right\}}\nonumber\\
& \times \log\left(\frac{M^2}{m_a^2}\right) + \frac{2 m_t^2 \left(m_a^4 - m_a^2 q^2 (1 + \xi)^2 + 2 q^4 \xi (1 + \xi)^2\right) \log\left(\frac{M^2}{m_a^2}\right)}{q^2 (q^2-4 m_t^2) (1 + \xi) \left(m_t^2 + q^2 \xi (1 + \xi)\right) \left(M^2 + \xi \left(m_a^2 - q^2 (1 + \xi)\right)\right)} \nonumber\\
& + \frac{\xi \left(2 m_a^4 - m_a^2 q^2 (1 + \xi) + q^4 (1 + \xi)^2 (-1 + 2 \xi)\right) }{(q^2-4 m_t^2) (m_t^2 + q^2 \xi (1 + \xi)) (M^2 + \xi (m_a^2 - q^2 (1 + \xi)))} \log\left(\frac{M^2}{m_a^2}\right) \nonumber\\
& +\frac{4 m_a^2 \left(m_a^2 - q^2\right) }{q^2 \left(q^2-4 m_t^2\right) \left(M^2 + \xi \left(m_a^2 - q^2 (1 + \xi)\right)\right)} \log\left(\frac{m_a^2}{m_a^2 - q^2}\right)\nonumber\\
& - \frac{4 q^2 \left(q^2-4 m_t^2\right) \xi^2 \left\{M^4 + q^4 \xi^3 (1 + \xi) - M^2 q^2 \xi (1 + 2 \xi)\right\}}{\left\{M^2 + \xi \left(m_a^2 - q^2 (1 + \xi)\right)\right\} \Lambda\left[m_t^2, q^2 \xi^2, m_t^2 + q^2 \xi (1 + \xi)\right]^2} \log\left(\frac{M^2}{M^2 - q^2 \xi^2}\right) \nonumber\\
& - \frac{2 m_a^2 \left(m_a^2 - q^2\right) }{\left(4 m_t^2 - q^2\right) \left\{M^2 + \xi \left(m_a^2 - q^2 (1 + \xi)\right)\right\}} f_2\left[m_t^2, m_t^2, q^2, m_a, m_t, 0\right] \nonumber\\
& + \frac{2 q^4 \left(q^2-4 m_t^2\right) \xi^3 \left(M^4 + q^4 \xi^3 (1 + \xi) - M^2 q^2 \xi (1 + 2 \xi)\right) }{\left\{M^2 + \xi \left(m_a^2 - q^2 (1 + \xi)\right)\right\} \Lambda\left[m_t^2, q^2 \xi^2, m_t^2 + q^2 \xi (1 + \xi)\right]^2}\nonumber\\
& \times f_2\left[m_t^2, q^2 \xi^2, m_t^2 + q^2 \xi (1 + \xi), m_t, 0, M\right] \nonumber\\
&- \frac{2 m_a^2 q^4 \left(q^2-4 m_t^2\right) (1 + \xi)^4 \left\{M^2 - q^2 (1 + \xi)^2 + m_a^2 (1 + 2 \xi)\right\}}{\left\{M^2 + \xi \left(m_a^2 - q^2 (1 + \xi)\right)\right\} \Lambda\left[m_t^2, q^2 (1 + \xi)^2, m_t^2 + q^2 \xi (1 + \xi)\right]^2}\nonumber\\
& \times f_2\left[m_t^2, q^2 (1 + \xi)^2, m_t^2 + q^2 \xi (1 + \xi), m_t, m_a, M\right]\Bigg]~,
\end{align}

\begin{align}
\mathcal{F}_6(q^2) =& -\frac{2y_{att}^2 g_s^3 m_t}{(4\pi)^4} c_t \tilde{c}_t\int_{0}^1 dx\int_{0}^{1-x}dy\,\left[\frac{1}{y(y-1)}\right]\,\nonumber\\
&\times \Bigg[\frac{4 m_t^2 (1 + \xi) f_1[m_t^2, m_t, m_a]}{\Lambda[m_t^2, 
 q^2 (1 + \xi)^2, m_t^2 + q^2 \xi (1 + \xi)]} + \frac{2 \left\{M^2 + \xi \left(m_a^2 - q^2 (1 + \xi)\right)\right\} }{q^2 (q^2-4 m_t^2) \xi (1 + \xi)} \log\left(\frac{m_t^2}{m_a^2}\right)\nonumber\\
 & + \frac{2 q^2 (1 + \xi)^2 \left(-M^2 + q^2 (1 + \xi)^2 - m_a^2 (1 + 2 \xi)\right) f_1\left[q^2 (1 + \xi)^2, M, m_a\right]}{\left(M^2 + \xi \left(m_a^2 - q^2 (1 + \xi)\right)\right) \Lambda\left[m_t^2, q^2 (1 + \xi)^2, m_t^2 + q^2 \xi (1 + \xi)\right]}\nonumber\\
 & +\frac{4 q^2 \left(q^2-4 m_t^2\right) \xi (1 + \xi) \left(m_t^2 + q^2 \xi (1 + \xi)\right) f_1\left[m_t^2 + q^2 \xi (1 + \xi), m_t, M\right]}{\Lambda\left[m_t^2, q^2 \xi^2, m_t^2 + q^2 \xi (1 + \xi)\right] \Lambda\left[m_t^2, q^2 (1 + \xi)^2, m_t^2 + q^2 \xi (1 + \xi)\right]}\nonumber\\
 & - \frac{M^4 (2 + \xi) - 2 M^2 \xi \left(-m_a^2 + q^2 (1 + \xi)^2\right) }{q^2 \left(q^2-4 m_t^2\right) \xi (1 + \xi)^2 \left\{M^2 + \xi \left(m_a^2 - q^2 (1 + \xi)\right)\right\}}\log\left(\frac{M^2}{m_a^2}\right) \nonumber\\
 &- \frac{\xi \left(-2 m_a^2 q^2 (1 + \xi)^2 + m_a^4 (1 + 2 \xi) + q^4 (1 + \xi)^2 (1 + \xi^2)\right) }{q^2 (q^2-4 m_t^2) \xi (1 + \xi)^2 \left(M^2 + \xi (m_a^2 - q^2 (1 + \xi))\right)}\log\left(\frac{M^2}{m_a^2}\right)\nonumber\\
 & -\frac{2 \left(m_a^2 - q^2\right)^2 }{q^2 \left(q^2-4 m_t^2\right) \left(M^2 + \xi \left(m_a^2 - q^2 (1 + \xi)\right)\right)} \log\left(\frac{m_a^2}{m_a^2 - q^2}\right) \nonumber\\
 &+\frac{2 \left(M^2 - q^2 \xi^2\right)^2 }{\left\{M^2 + \xi \left(m_a^2 - q^2 (1 + \xi)\right)\right\} \Lambda[m_t^2, q^2 \xi^2, m_t^2 + q^2 \xi (1 + \xi)]} \log\left(\frac{M^2}{M^2 - q^2 \xi^2}\right) \nonumber\\
 & - \frac{4 m_t^2 (m_a^2 - q^2)^2 }{q^2 (q^2-4 m_t^2) \left(M^2 + \xi (m_a^2 - q^2 (1 + \xi))\right)} f_2[m_t^2, m_t^2, q^2, m_a, m_t, 0] \nonumber\\
 & - \frac{4 m_t^2 \xi \left(M^2 - q^2 \xi^2\right)^2 f_2\left[m_t^2, q^2 \xi^2, m_t^2 + q^2 \xi (1 + \xi), m_t, 0, M\right]}{\left\{M^2 + \xi \left(m_a^2 - q^2 (1 + \xi)\right)\right\} \Lambda\left[m_t^2, q^2 \xi^2, m_t^2 + q^2 \xi (1 + \xi)\right]} \nonumber\\
 & + \frac{4 m_a^2 q^2 (1 + \xi)^2 f_2[m_t^2, q^2 (1 + \xi)^2, 
  m_t^2 + q^2 \xi (1 + \xi), m_t, m_a, M]}{\Lambda[m_t^2, 
 q^2 (1 + \xi)^2, m_t^2 + q^2 \xi (1 + \xi)]} \nonumber\\
 & + \frac{4 m_t^2 (1 + \xi) \left\{M^4 + \left(m_a^2 - q^2 (1 + \xi)^2\right)^2 - 2 M^2 \left(m_a^2 + q^2 (1 + \xi)^2\right)\right\} }{\left\{M^2 + \xi \left(m_a^2 - q^2 (1 + \xi)\right)\right\} \Lambda\left[m_t^2, q^2 (1 + \xi)^2, m_t^2 + q^2 \xi (1 + \xi)\right]}\nonumber\\
 & \times f_2\left[m_t^2, q^2 (1 + \xi)^2, m_t^2 + q^2 \xi (1 + \xi), m_t, m_a, M\right]\Bigg]~,
\end{align}
where we have used the following definitions:
\begin{align}
    &\xi = \frac{x}{y-1}~, \hspace{1cm} M = \left[ \frac{m_t^2}{y(1-y)} + \frac{q^2 x (x+y-1)}{y(1-y)^2}\right]^\frac{1}{2}~,\nonumber\\
    &\Lambda[a,b,c] = a^2 + b^2 + c^2 -2ab - 2bc -2ca~.
\end{align}

The function $f_1\left[m_1, m_2, m_3\right]$ is given by

\begin{align}
    f_1\left[m_1, m_2, m_3\right] = \frac{\Lambda\big[m_1, m_2^2, m_3^2\big]}{m_1}  \log\left(\frac{m_2^2+m_3^2-m_1 + \Lambda\big[m_1, m_2^2, m_3^2\big]}{2 m_2 m_3}\right)~.
\end{align}

The function $f_2\left[s_1, s_2, s_3, m_1, m_2, m_3\right]$ is given by

\begin{align}
f_2\big[s_1, s_2,& s_3, m_1, m_2, m_3\big] = \frac{1}{\Lambda[s_1, s_2, s_3]}\nonumber\\
&\Biggl\{-\mathrm{Li}_2\Bigg[\frac{\Phi(s_1, s_2, s_3, m_1, m_2, m_3) - \left(m_1^2 - m_2^2 + s_1\right) \sqrt{\Lambda[s_1, s_2, s_3]}}{\Phi(s_1, s_2, s_3, m_1, m_2, m_3) - \sqrt{\Lambda[m_1^2, m_2^2, s_1] \Lambda[s_1, s_2, s_3]}}\nonumber\\
&-i\epsilon s_1 \left(g_1(s_1,s_2,s_3) m_1^2 - g_2(s_1,s_2,s_3) m_2^2 + g_1(s_3,s_2,s_1) s_1 + 2 m_3^2 s_1\right)\Bigg]\nonumber\\
& + \mathrm{Li}_2\Bigg[\frac{\Phi(s_1, s_2, s_3, m_1, m_2, m_3) + \left(-m_1^2 + m_2^2 + s_1\right) \sqrt{\Lambda[s_1, s_2, s_3]}}{\Phi(s_1, s_2, s_3, m_1, m_2, m_3) - \sqrt{\Lambda[m_1^2, m_2^2, s_1] \Lambda[s_1, s_2, s_3]}}\nonumber\\
&-i\epsilon s_1 \left(g_1(s_1, s_2, s_3) m_1^2 - g_2(s_1,s_2,s_3) m_2^2 - g_2(s_2,s_1,s_3) s_1 + 2 m_3^2 s_1\right)\Bigg]\nonumber\\
&-\mathrm{Li}_2\Bigg[\frac{\Phi(s_1, s_2, s_3, m_1, m_2, m_3) - \left(m_1^2 - m_2^2 + s_1\right) \sqrt{\Lambda[s_1, s_2, s_3]}}{\Phi(s_1, s_2, s_3, m_1, m_2, m_3) + \sqrt{\Lambda[m_1^2, m_2^2, s_1] \Lambda[s_1, s_2, s_3]}} \nonumber\\
&+i\epsilon s_1 \left(g_1(s_1,s_2,s_3) m_1^2 - g_2(s_1,s_2,s_3) m_2^2 + g_1(s_3,s_2,s_1) s_1 + 2 m_3^2 s_1\right)\Bigg]\nonumber\\
&+\mathrm{Li}_2\Bigg[\frac{\Phi(s_1, s_2, s_3, m_1, m_2, m_3) + \left(-m_1^2 + m_2^2 + s_1\right) \sqrt{\Lambda[s_1, s_2, s_3]}}{\Phi(s_1, s_2, s_3, m_1, m_2, m_3) + \sqrt{\Lambda[m_1^2, m_2^2, s_1] \Lambda[s_1, s_2, s_3]}}\nonumber\\
&+i\epsilon s_1 \left(g_1(s_1,s_2,s_3) m_1^2 - g_2(s_1,s_2,s_3) m_2^2 - g_2(s_2,s_1,s_3) s_1 + 2 m_3^2 s_1\right)\Bigg]\nonumber\\
&-\mathrm{Li}_2\Bigg[\frac{\Phi(s_3, s_2, s_1, m_1, m_3, m_2)  + \left(m_1^2 - m_3^2 - s_3\right) \sqrt{\Lambda[s_1, s_2, s_3]}}{\Phi(s_3, s_2, s_1, m_1, m_3, m_2) + \sqrt{\Lambda[m_1^2, m_3^2, s_3] \Lambda[s_1, s_2, s_3]}}\nonumber\\
&+i\epsilon s_3 \left(-g_2(s_2,s_1,s_3) m_1^2 - g_2(s_1,s_2,s_3) m_3^2 + g_1(s_1,s_2,s_3) s_3 + 2 m_2^2 s_3\right)\Bigg]\nonumber\\
&+ \mathrm{Li}_2\Bigg[\frac{\Phi(s_3, s_2, s_1, m_1, m_3, m_2) + \left(m_1^2 - m_3^2 + s_3\right) \sqrt{\Lambda[s_1, s_2, s_3]}}{\Phi(s_3, s_2, s_1, m_1, m_3, m_2) + \sqrt{\Lambda[m_1^2, m_3^2, s_3] \Lambda[s_1, s_2, s_3]}}\nonumber\\
&+i\epsilon s_3 \left(-g_2(s_2,s_1,s_3) m_1^2 - g_2(s_1,s_2,s_3) m_3^2 - g_2(s_1,s_3,s_2) s_3 + 2 m_2^2 s_3\right)\Bigg]\nonumber\\
&+\mathrm{Li}_2\Bigg[\frac{-\Phi(s_3, s_2, s_1, m_1, m_3, m_2) - \left(m_1^2 - m_3^2 + s_3\right) \sqrt{\Lambda[s_1, s_2, s_3]}}{-\Phi(s_3, s_2, s_1, m_1, m_3, m_2) + \sqrt{\Lambda[m_1^2, m_3^2, s_3] \Lambda[s_1, s_2, s_3]}}\nonumber\\
&-i\epsilon s_3 \left(-g_2(s_2,s_1,s_3) m_1^2 - g_2(s_1,s_2,s_3) m_3^2 - g_2(s_1,s_3,s_2) s_3 + 2 m_2^2 s_3\right)\Bigg]\nonumber\\
&-\mathrm{Li}_2\Bigg[\frac{-\Phi(s_3, s_2, s_1, m_1, m_3, m_2) + \left(-m_1^2 + m_3^2 + s_3\right) \sqrt{\Lambda[s_1, s_2, s_3]}}{-\Phi(s_3, s_2, s_1, m_1, m_3, m_2) + \sqrt{\Lambda[m_1^2, m_3^2, s_3] \Lambda[s_1, s_2, s_3]}}\nonumber\\
&-i\epsilon s_3 \left(-g_2(s_2,s_1,s_3) m_1^2 - g_2(s_1,s_2,s_3) m_3^2 + g_1(s_1,s_2,s_3) s_3 + 2 m_2^2 s_3\right)\Bigg]\nonumber\\
&- \mathrm{Li}_2\Bigg[\frac{\Phi(s_2, s_1, s_3, m_3, m_2, m_1) - \left(m_2^2 - m_3^2 + s_2\right) \sqrt{\Lambda[s_1, s_2, s_3]}}{\Phi(s_2, s_1, s_3, m_3, m_2, m_1) + \sqrt{\Lambda[m_2^2, m_3^2, s_2] \Lambda[s_1, s_2, s_3]}}\nonumber\\
&+i\epsilon s_2 \left(g_1(s_3,s_2,s_1) m_2^2 - g_2(s_1,s_2,s_3) m_3^2 - g_2(s_1,s_2,s_3) s_2 + 2 m_1^2 s_2\right)\Bigg]\nonumber\\
&+ \mathrm{Li}_2\Bigg[\frac{\Phi(s_2, s_1, s_3, m_3, m_2, m_1) + \left(-m_2^2 + m_3^2 + s_2\right) \sqrt{\Lambda[s_1, s_2, s_3]}}{\Phi(s_2, s_1, s_3, m_3, m_2, m_1) + \sqrt{\Lambda[m_2^2, m_3^2, s_2] \Lambda[s_1, s_2, s_3]}}\nonumber\\
&+i\epsilon s_2 \left(g_1(s_3,s_2,s_1) m_2^2 - g_2(s_1,s_2,s_3) m_3^2 - g_2(s_1,s_2,s_3) s_2 + 2 m_1^2 s_2\right)\Bigg]\nonumber\\
&+ \mathrm{Li}_2\Bigg[\frac{-\Phi(s_2, s_3, s_1, m_2, m_3, m_1) + \left(m_2^2 - m_3^2 - s_2\right) \sqrt{\Lambda[s_1, s_2, s_3]}}{-\Phi(s_2, s_3, s_1, m_2, m_3, m_1) + \sqrt{\Lambda[m_2^2, m_3^2, s_2] \Lambda[s_1, s_2, s_3]}}\nonumber\\
&-i\epsilon s_2 \left(g_1(s_3,s_2,s_1) m_2^2 - g_2(s_1,s_2,s_3) m_3^2 - g_2(s_1,s_2,s_3) s_2 + 2 m_1^2 s_2\right)\Bigg]\nonumber\\
&-\mathrm{Li}_2\Bigg[\frac{-\Phi(s_2, s_3, s_1, m_2, m_3, m_1) + \left(m_2^2 - m_3^2 + s_2\right) \sqrt{\Lambda[s_1, s_2, s_3]}}{-\Phi(s_2, s_3, s_1, m_2, m_3, m_1) + \sqrt{\Lambda[m_2^2, m_3^2, s_2] \Lambda[s_1, s_2, s_3]}}\nonumber\\
&-i\epsilon s_2 \left(g_1(s_3,s_2,s_1) m_2^2 - g_2(s_1,s_2,s_3) m_3^2 - g_2(s_1,s_2,s_3) s_2 + 2 m_1^2 s_2\right)\Bigg]\Biggr\}~,
\end{align}
where $\Phi(s_1,s_2,s_3, m_1,m_2,m_3) = m_1^2 (s_1 + s_2 - s_3) + m_2^2 (s_1 - s_2 + s_3) + 
 s_1 (-2 m_3^2 - s_1 + s_2 + s_3)$, $g_1(s_1, s_2, s_3) = -s_1 - s_2 + s_3 + \sqrt{\Lambda[s_1,s_2,s_3]}$, and $g_2(s_1, s_2, s_3) = s_1 - s_2 + s_3 + \sqrt{\Lambda[s_1,s_2,s_3]}$. 
 The dilogarithm is defined as follows:
 
\begin{align}
    \mathrm{Li}_2 (z) = -\int_{0}^{z} \frac{\log(1-t)}{t} dt \ \text{ for } \ z\in \mathbb{C}~.
\end{align}




\bibliographystyle{CitationStyle}
\bibliography{myrefs}
\end{document}